\documentclass[sigconf, nonacm]{acmart}
\usepackage{tcolorbox}
\usepackage{booktabs} 
\usepackage{multirow}
\usepackage{siunitx}
\usepackage{enumitem}
\usepackage{adjustbox}
\usepackage{placeins}
\newcommand{\lsm}{\textsc{CascadeLSM}}

\begin{document}
\title{USDB: Scaling Storage for Next-Generation Recommender Systems at ByteDance}

\author{
    Xiaoran Zhu$^{*}$,
    Jieming Cai$^{*}$,
    Weiping Yu$^{\dagger}$,
    Xi Chen,
    Shu Cai,
    Jiahao Li,
    Yifan Xiao,
        Yan Deng,
    Yisha Zhou,
        Yu Lin,
        Shiru Ren,
    Siqiang Luo$^{\ddagger\dagger}$,
    Peng Xu,
    Junjun Chai
}
\affiliation{
  ByteDance Inc.;
   $^\ddagger$Nanyang Technological University
}

\begin{abstract}
Recommendation models consume increasingly long, attribute-rich user histories, making small continuous updates and repeated large reads a central storage asymmetry. We present USDB, ByteDance's unified sequence storage practice for streaming and batch workloads. \lsm{} time-based compaction stages materialize application-specific histories; differentiated replicas share persistent files; partial dumps reconstruct inference inputs from captured recent state and versioned history; and a hybrid columnar format accelerates projected reads. These designs reuse maintenance results across ingestion, inference, and training. Production migration records show 85.8\% lower mean write latency, while normalized deployment accounting shows 42.1\% lower infrastructure cost. Controlled comparisons with an Optimized RocksDB sequence baseline show faster long-history reads across projections and client loads. Aggregation pushdown sustains 8.5$\times$ the throughput of ordinary reads under a constrained receive budget. Production reconstruction and direct-file training results demonstrate how shared sequence maintenance supports fresh serving and repeated historical access.
\end{abstract}

\maketitle

\renewcommand\thefootnote{}
\footnotetext{\noindent * equal contribution.}
\footnotetext{\noindent $\dagger$ corresponding authors, siqiang.luo@ntu.edu.sg and weiping.yu@bytedance.com.}
\pagestyle{plain}

\raggedbottom
\setlength{\emergencystretch}{2em}
\setcounter{topnumber}{1}
\setcounter{dbltopnumber}{1}
\setcounter{totalnumber}{1}
\section{Introduction}
Recommendation systems~\cite{ricci2021recommender,resnick1997recommender,bobadilla2013recommender}, widely used in online content platforms and e-commerce, personalize billions of user experiences daily. At ByteDance, recommendation powers a diverse portfolio of products including videos in TikTok, Douyin, live streaming, and e-commerce platforms, serving billions of users worldwide. Over the past decade, the field has evolved from traditional collaborative filtering to Deep Learning Recommendation Models (DLRMs)~\cite{mudigere2022software,cheng2016wide,zhou2018deep,covington2016deep}, and most recently, to Large Recommendation Models (LRMs)~\cite{zhai2024actions,chai2025longer,zhao2024recommender} that fundamentally reshape how recommendations are generated. Unlike DLRMs, which rely on heterogeneous feature engineering and do not scale efficiently with additional compute on more feature interactions or deeper layers~\cite{zhai2024actions, wang2025scaling}, LRMs adopt sequential transduction architectures that directly model user behavior sequences, exhibiting compelling scaling laws~\cite{kaplan2020scaling, zhai2024actions, guan2025make,chai2025longer}: model quality improves predictably with longer user behavior sequences, denser computation in attention layers, and increased model depth.

\begin{figure}[t]
    \centering
    \includegraphics[width=\columnwidth]{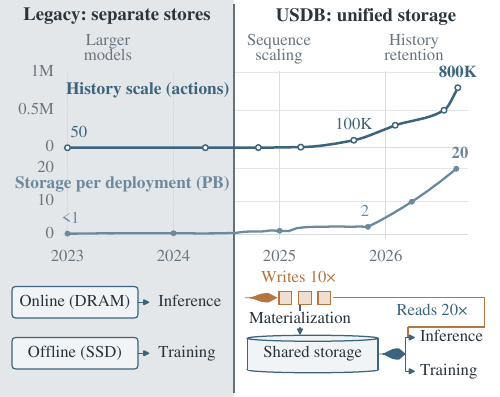}
    \caption{Recommendation-storage evolution through September 2026. Representative history and deployment-scale estimates accompany the storage organizations; traffic factors summarize approximate operational growth.}
    \label{fig:traffic}
\end{figure}

These scaling trends increase the history retained and retrieved by recommendation infrastructure. At ByteDance, as shown in Figure~\ref{fig:traffic}, the sequence length consumed by recommendation models has increased 2,000-fold from 2023 to 2025 (50 to 100,000), with representative retained-history configurations reaching 800,000 actions in September 2026, while the number of feature columns has grown from dozens to hundreds. Besides this algorithmic shift, the explosive growth in user base and engagement frequency has also amplified data volumes---Figure~\ref{fig:traffic} summarizes the corresponding growth in deployment storage and the transition from separate stores to shared sequence storage. The plotted stages are representative internal estimates across applications; deployment capacity combines historical estimates with later scale anchors.

\begin{figure}[t]
    \centering
    \includegraphics[width=\columnwidth]{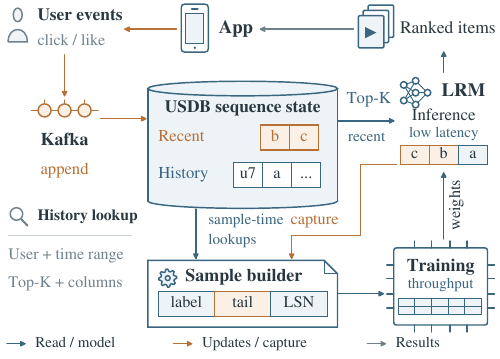}
    \caption{Recommendation access. Inference reads recent Top-$K$ histories; training repeats sample-time lookups. Both filter by user/time and select columns. Letters mark actions.}
    \label{fig:dataflow}
\end{figure}

Supporting these large-scale recommendation workloads requires robust data infrastructure capable of simultaneously serving both streaming (real-time) and batch (historical) data access patterns as shown in Figure~\ref{fig:dataflow}. For streaming inference, the system must provide low-latency point lookups to retrieve recent user interactions. For batch training, it must deliver high-throughput historical lookups and scans over massive datasets. Feature stores~\cite{de2024hopsworks,liu2023optimizing,orr2021managing} have emerged as an important component in ML infrastructure, providing feature management, discovery, and serving capabilities at the application layer. However, the underlying storage engine that powers these systems remains a critical yet under-explored challenge, particularly when handling petabyte-scale \textbf{user behavior sequences}~\cite{lai2025modeling,he2023survey} with the dual access patterns described above. This dual requirement naturally suggests a Lambda architecture~\cite{warren2015big,kreps2014questioning} that processes real-time and batch data through separate pipelines, ultimately merging them for downstream consumption by training and inference systems. On the storage engine side, key-value storage using LSM-trees naturally aligns with our target due to its write-optimized structure~\cite{dayan2017monkey,dayan2018dostoevsky}. Combining these two techniques yields a promising storage architecture for modern recommendation systems in the large model era.

\textbf{Challenges.} Scaling this architecture to billion-user platforms with petabyte-scale data introduces four fundamental challenges. First, existing approaches aggregate user sequences in memory using a read-before-write pattern; as streaming volumes exceed memory capacity and spill to SSDs, this transforms each small write into an SSD-bound bottleneck with severe amplification (\S~\ref{sec:ingest}). Second, LSM-tree compaction is CPU-intensive while batch storage demands large disk capacity, and coupling both on the same nodes drives up infrastructure costs and complicates capacity planning (\S~\ref{sec:imbalance}). Third, maintaining consistency between training and inference becomes increasingly difficult at scale: out-of-order arrivals and delayed feedback signals cause the sequences observed during inference to diverge from those used in training, and LRMs are highly sensitive to such skew (\S~\ref{sec:inconsist}). Finally, the conventional row-oriented SST format forces training workloads to deserialize entire records when scanning specific feature columns, wasting I/O bandwidth and yielding suboptimal compression ratios (\S~\ref{sec:inefficiency}).

\textbf{Our System.} To address these challenges, we developed USDB, a universal user behavior sequence data backend for ByteDance's entire recommendation system. This paper shares our industrial experience in building and operating a production-level storage system that serves one of the world's largest recommendation platforms, involving billions of users, petabyte-scale retained histories, and large distributed training and serving fleets. Our industrial experience connects storage architecture, consistency mechanisms, and data formats within one sequence lifecycle.

The contributions of this paper are as follows:

\noindent\textbf{Architecture Contribution.}
We organize streaming ingestion, historical materialization, inference, and training around a shared sequence lifecycle. Application-specific aggregation produces reusable history; differentiated replicas publish and reuse its persistent files; recorded visibility connects inference inputs to later training reads; and a common indexed columnar format supports both consumers. This coordination turns the four challenges into the four corresponding mechanisms below.

\noindent\textbf{System Design.} USDB (\S~\ref{sec:arch}) integrates four components:
\begin{itemize}[leftmargin=*, align=left,topsep=0pt,partopsep=0pt]
     \item We propose \lsm{}, a unified time-tiered LSM-tree that organizes streaming and batch data within a single structure. Through time-based data organization and differentiated compaction strategies, \lsm{} eliminates the read-before-write pattern that plagued previous systems, reducing write latency by 85\% while maintaining read latency (\S~\ref{sec:lsm}).

    \item We design a differentiated multi-replica architecture over disaggregated storage-compute architecture, comprising read-write, read-heavy, and read-only replicas each tailored to its workload. Lightweight synchronization ensures replica consistency without redundant compaction, achieving a 42\% reduction in total infrastructure cost.

    \item We develop a {sequence number-based versioning} mechanism to maintain consistency between training and inference workloads. Combined with our partial dump mechanism, this reconstructs earlier inference inputs while replacing repeated full-sequence dumps with a captured recent fragment and reusable history (\S~\ref{sec:mvcc}).

    \item We design a {hybrid columnar encoding} format with adaptive encoding selection that serves both access patterns: columnar scans for training throughput and row-level indexing for inference latency. Our approach achieves 4.6--5.5$\times$ faster decompression than ZSTD in the codec evaluation, with encoded sizes between LZ4 and ZSTD (\S~\ref{sec:encoding}).
\end{itemize}

\noindent\textbf{Production Experience.} Deployed since 2024 across Douyin, TikTok, and their live streaming platforms, USDB reduces mean write latency by 85.8\%, mean CPU utilization by 72.6\%, and normalized infrastructure cost by 42.1\%. A 374-run study against \emph{Optimized RocksDB} complements the production evaluation: repeated 10K/100K reads achieve 1.79$\times$/1.75$\times$ throughput, and 100K client-read jobs complete 1.62--2.52$\times$ faster (\S~\ref{sec:eval}).

\section{Background}
We introduce the sequence workloads, processing architecture, and storage primitives underlying USDB.

\subsection{Long User Histories}
LRMs use sequential transduction to model ordered user interactions, capture interests, and generate recommendations. Each interaction records a user, timestamp, and multidimensional features such as item identifiers, categories, and engagement signals. At ByteDance, a user's history can contain tens of thousands of records, with petabytes in aggregate. Online inference retrieves a user's sequence through a low-latency lookup keyed by uid to construct model input in real time. Training retrieves histories for many labeled samples through indexed lookups or bulk historical scans. These consumers impose different physical requirements: point lookups favor per-user locality and fast random access, whereas scans favor columnar efficiency and high I/O throughput (Table~\ref{tab:patterns}). USDB combines these access patterns with application-specific aggregation and reconstruction of the history visible to an earlier inference request.

\begin{table}[t]
\small\centering
\setlength{\tabcolsep}{3pt}
\caption{Consumers of a shared logical user history.}
\label{tab:patterns}
\begin{tabular}{@{}>{\raggedright\arraybackslash}p{.19\columnwidth}>{\raggedright\arraybackslash}p{.36\columnwidth}>{\raggedright\arraybackslash}p{.38\columnwidth}@{}}
\toprule
Consumer & Access & Dominant requirement \\\midrule
Inference & User, time range, Top-$K$, columns & Fresh observations; request latency \\
Training & Historical lookups for many samples & Aggregate reads; original inference state \\
Maintenance & Increments, materialization, backfill & Reusable history; bounded pending work \\
\bottomrule
\end{tabular}
\vspace{-2mm}
\end{table}

\subsection{Lambda Architecture}
Lambda architecture~\cite{warren2015big} separates a speed layer that processes streams for low-latency serving from a batch layer that periodically recomputes views over complete history; a serving layer combines their results. Recommendation deployments use recent interactions for real-time inference and full histories to generate training datasets. Traditional implementations maintain separate stores, duplicating data and maintenance while complicating consistency and operation~\cite{kreps2014questioning}. USDB unifies their storage while preserving both consumers' latency and throughput needs.

\subsection{Log-Structured Merge-Tree}
Log-Structured Merge-Tree (LSM-tree)~\cite{o1996log} is a write-optimized data structure widely adopted in modern key-value stores. Writes are first appended to a Write-Ahead Log (WAL) for durability and buffered in memory; when the buffer fills, it is flushed as an immutable Sorted String Table (SST) file organized into multiple levels, with compaction periodically merging overlapping files to reclaim space and improve read performance. Since SST files are immutable, both compaction reads and writes exhibit sequential access patterns that align well with distributed file systems optimized for append-only operations. This design naturally suits our requirements: its write-optimized structure handles high-velocity user interaction streams, while compaction supports time-based data organization and aggregation.

\subsection{Storage Formats and Encoding}
Row-oriented storage formats store each key-value pair contiguously, optimizing for point lookups but performing poorly when scanning specific columns across many rows, a common pattern in training sample generation. Columnar formats instead store each column contiguously, enabling efficient scans and higher compression ratios, but incur overhead for point lookups and small updates. Since our workload requires both scanning specific feature columns across millions of user sequences for training and retrieving complete user records for inference, we adopt a hybrid design that adapts storage formats and encodings to access patterns.

\paragraph{Sequence selection and visibility.}
\label{sec:querycontract}
A history request specifies a user, event-time interval, maximum sequence length (Top-$K$), and selected columns. Reconstruction also uses the ingestion-version boundary recorded by inference. The query first selects records visible at that boundary, then orders them by descending event time, retains the requested number, and projects the selected columns.
The application defines repeated-interaction, attribute-merge, and tie rules. Event time selects relevant behavior; ingestion version identifies what had arrived. Visibility selection precedes ordering and Top-$K$, so a later-arriving event with an old timestamp cannot displace an event that inference observed. A partial dump carries the original selection parameters and a recent fragment; training retrieves a version-compatible older prefix from retained history.

\begin{figure*}[t]
\centering
\includegraphics[width=\textwidth]{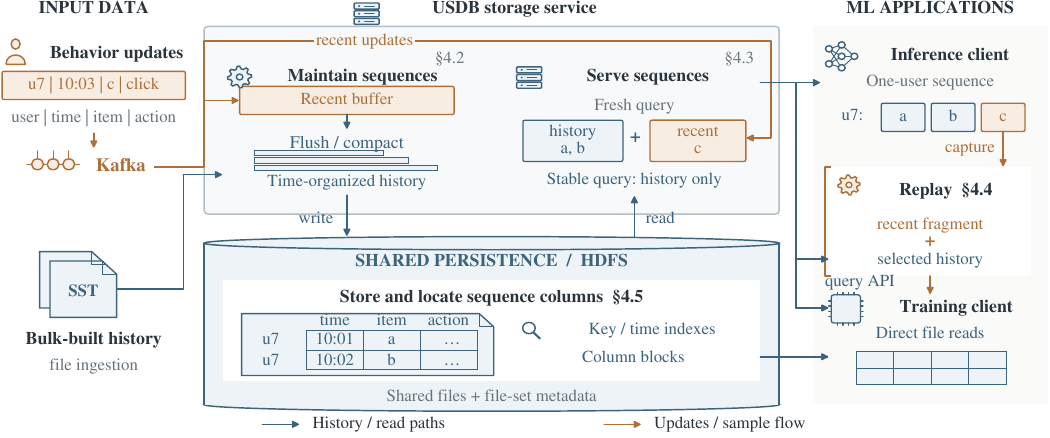}
\caption{USDB architecture: streaming updates and bulk-built history share maintenance, persistent files, and sequence access. Fresh queries combine recent state with history; training can query or read shared files and reconstruct earlier inference inputs from partial dumps.}
\label{fig:arc}
\end{figure*}

\begin{figure*}[t]

    \includegraphics[width=\textwidth]{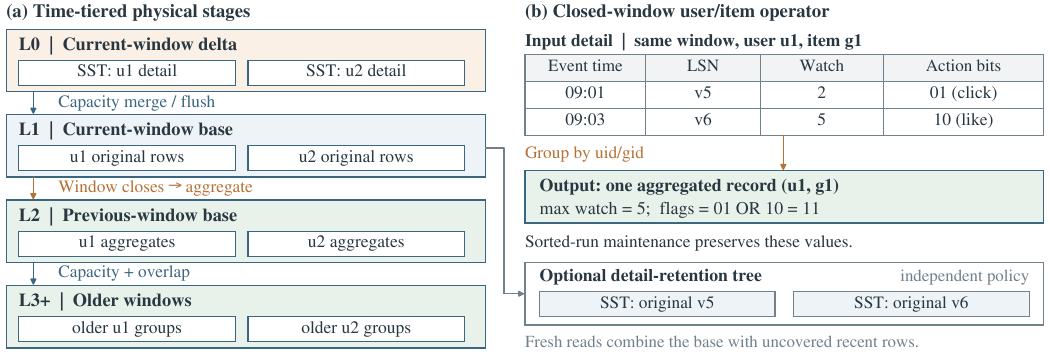}
    \caption{\lsm{} stages and aggregation. Current-window deltas and bases retain detail; closing a window aggregates by user/item. The example combines watch-time maxima and action-flag union. Older compactions preserve aggregates, while a separately retained detail tree keeps original rows.}
    \label{fig:cascade}
\end{figure*}

\begin{figure*}[t]
    \centering
    \includegraphics[width=\textwidth]{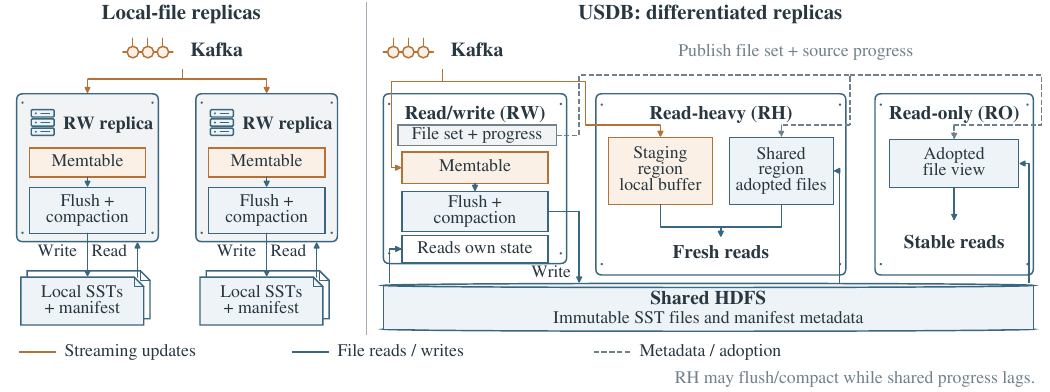}
    \caption{Replica organization before and after disaggregation: local files versus differentiated roles over shared storage.}
    \label{fig:sep}
\end{figure*}

\begin{figure*}[t]
    \centering
    \includegraphics[width=\textwidth]{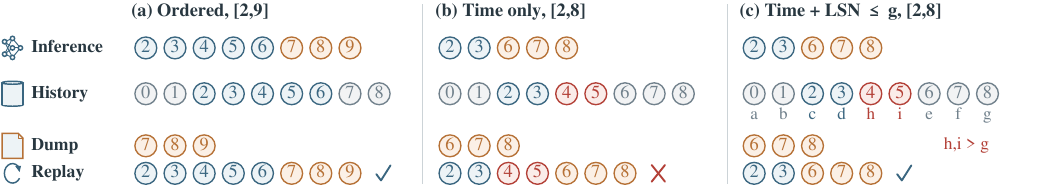}
    \caption{Partial-dump replay. Circles show event times; letters below show LSNs. Gray: skipped history; orange: captured tail; red: late arrivals. Time-only replay adds 4 and 5; MVCC excludes $h,i>g$. Ranges are inclusive; Top-$K$ is nonbinding.}
    \label{fig:dump}
\end{figure*}

\section{Motivation and Challenges}
\label{sec:challenge}
ByteDance's early recommendation systems partitioned user sequences across distributed memory stores for low-latency inference. As the user base grew to billions and histories deepened, petabyte-scale sequence data exceeded memory capacity. Persistent storage must now support millisecond-level inference lookups and high-throughput historical training reads over a continuously growing dataset. Unifying these demands cost-effectively exposes four challenges.

\subsection{Streaming Ingestion}
\label{sec:ingest}
In our early sequence storage implementations with moderate data volumes, we maintained streaming data entirely in memory and batch data on local SSDs, operating two independent databases. To optimize read performance for streaming queries, we aggregated user behavior sequences in memory using a key-key-value (KKV) format organized as \texttt{date$\rightarrow$uid$\rightarrow$actions}. This design enables a single point lookup to retrieve all interactions for a given user on a specific date. However, this aggregation strategy requires reading the entire existing sequence before each write operation to perform the merge, resulting in significant write amplification. While acceptable when all data resides in memory, this approach becomes untenable as write volumes grow.

As user activity grew, streaming data spilled to SSD, turning this amplification into an I/O bottleneck. Appending one interaction to a sequence with thousands of actions requires reading prior data from SSD, merging in memory, and writing the complete sequence back. This amplifies small updates by orders of magnitude.

Separate streaming and batch stores also duplicate data and maintenance. Periodic transfers create consistency windows and complicate recovery. We need a unified structure that preserves efficient streaming reads while eliminating read-before-write beyond memory capacity.

\subsection{Resource Imbalance}
\label{sec:imbalance}
LSM-tree compaction consumes CPU to merge and recompress SST files, while batch storage needs disk capacity for extended histories. Traditional architectures couple both requirements on the same nodes, tying compute and storage provisioning together.

This coupling creates several critical problems. First, different workload phases stress different resources. Streaming writes trigger compaction that consumes CPU, while batch reads demand sustained storage and network throughput, with CPU needs determined by decoding and sample construction. Provisioning for peak demand on both dimensions simultaneously leads to resource underutilization during normal operations. Second, capacity expansion is difficult: since each node's data resides locally, scaling requires copying all shard data to new nodes, creating substantial migration overhead and risking service disruption. This is particularly problematic given the rapid growth in data volumes, where clusters frequently require expansion.

The rigid CPU-to-storage ratio of container deployments exacerbates resource waste. For sequence feature workloads that are storage-heavy but compute-light, significant CPU cycles remain idle. Conversely, compaction-intensive periods may starve other operations of compute resources. Additionally, independent replicas using RocksDB~\cite{dong2021rocksdb,dong2023disaggregating} must each execute redundant compaction operations, multiplying computational waste across the cluster.

\subsection{Training-Inference Consistency}
\label{sec:inconsist}
Recommendation models exhibit high sensitivity to training-inference data discrepancies, a phenomenon known as training-serving skew. In production, we frequently observe scenarios where models show significant AUC improvements during offline training but exhibit metric degradation during online A/B testing. A primary contributing factor is inconsistency between the user behavior sequences observed during inference and those used during training.

Interactions arrive out of order because of network delays, processing lag, or delayed feedback. Long-form video engagement signals may arrive minutes or hours later. Training must also exclude information unavailable at prediction time to prevent feature leakage, requiring both temporal and visibility alignment.

The conventional approach to maintain consistency is to dump complete inference samples as snapshots for offline training. However, the combination of exponential growth in user activity and LRM's appetite for significantly longer and wider sequences has made the cost of full sample dumps prohibitive, across many overlapping samples. To address this, we designed a partial dump mechanism that captures only a subset of the inference state, specifically the most recent actions and relevant timestamps, then reconstructs complete training samples by merging with historical data from the offline cluster.

Reconstruction must preserve inference visibility: timestamp ranges alone cannot exclude late arrivals within the middle of a sequence. Network-induced sideinfo loss and failover retransmission further complicate the merge. A versioning mechanism must identify which records inference observed and filter incompatible arrivals during training.

\subsection{Storage Efficiency}
\label{sec:inefficiency}
Standard row-oriented SST formats optimize point lookups and small range scans. Complete key-value records are stored sequentially and indexed by key, serving inference efficiently but mismatching projected training reads.

Training sample generation exhibits a columnar access pattern: it scans specific feature columns (e.g., item IDs, timestamps, interaction types) across millions of user sequences while ignoring other columns. Row-oriented storage forces the system to read entire records from disk and then discard unused columns, wasting I/O bandwidth and cache space. With user sequences containing dozens of feature columns and training workloads accessing only a subset, this inefficiency multiplies across petabyte-scale datasets.

Additionally, row-oriented formats achieve suboptimal compression ratios. Compression algorithms exploit data locality and repetition, properties that manifest more strongly within individual columns than across heterogeneous column types in row-major layout. For instance, item ID columns exhibit strong categorical locality that enables aggressive compression through techniques like dictionary encoding, but mixing them with unrelated columns in row format dilutes these opportunities.

The challenge is to develop a storage format that maintains row-oriented efficiency for inference point lookups while enabling columnar scans and improved compression for training workloads, all within a unified storage system that avoids maintaining duplicate copies in different formats.

\section{Solution: USDB}
\label{sec:solution}
USDB addresses the four challenges through coordinated storage architecture, sequence maintenance, input reconstruction, and physical layout. Four coordinated mechanisms address the challenges: \lsm{} removes full-history rewriting from ingestion; differentiated replicas separate persistent capacity and maintenance from read serving; partial dumps with MVCC connect inference state to training; and hybrid columnar encoding accelerates selective historical access.

\subsection{Architecture}
\label{sec:arch}
Figure~\ref{fig:arc} shows USDB's four integrated components. At the storage-engine layer, \lsm{} organizes recent details and historical aggregates into time-based stages. Ingestion appends individual interactions; pluggable compaction constructs a user-local representation that inference and training can reuse. The hybrid columnar format combines per-user indexing with selective column access inside that representation.

At the system layer, read-write replicas consume interaction logs, materialize sequences, and publish immutable SST files to shared HDFS storage. Read-heavy replicas combine a recent local staging region with those files for fresh reads. Read-only replicas reuse persistent views for bulk historical consumption. A batch builder can produce compatible history for the same access interface. The organization shares the cost of aggregation, encoding, and persistence across consumers while allowing compute to be provisioned by role.

Partial dumps bridge online serving and later sample generation. Inference records the recent fragment and visibility metadata; training combines it with retained historical data. The metadata identifies the original selection and compatible history, connecting reconstruction to the same lifecycle that maintains and publishes the sequence. Each component addresses the corresponding challenge in Section~\ref{sec:challenge}.

\subsection{\lsm{}: Adaptive Compaction}
\label{sec:lsm}
The core storage component of USDB is CascadeLSM, which unifies streaming and batch data management through a common LSM maintenance organization with time-based tiers. Rather than maintaining separate storage systems for real-time serving and batch training as in traditional Lambda architectures, CascadeLSM leverages adaptive compaction strategies to serve both workloads from reusable sequence state. This design reduces the separate representations and pipeline synchronization while enabling workload-specific optimizations at each tier through pluggable compaction logic.

\subsubsection{Time-Tiered Organization}
We organize the LSM-tree into multiple logical levels based on data recency, as illustrated in Figure~\ref{fig:cascade}. Level 0 serves as the delta layer containing the most recent incremental data within the current processing-time window. Level 1 holds the accumulated base data for the current window. Level 2 stores the previous window's aggregated data, and deeper levels contain progressively older aggregated data. Each fixed processing-time interval $[w,w+\Delta)$ groups arrivals for materialization. The granularity is configurable: our production deployment typically uses daily windows to align with user activity and downstream training schedules; hourly or weekly windows are also supported.

The upper levels store detailed, fine-grained interaction records to support real-time inference queries. Each record represents a single user action with its complete attributes. The lower levels store aggregated data for two reasons. First, the sheer volume of detailed historical data makes full retention impractical at scale. Second, downstream training workloads often require pre-aggregated features such as deduplicated or merged user behaviors within session windows. Pre-computing these aggregations reduces both data transfer costs and computation burden on training pipelines. For scenarios that require access to fine-grained historical records beyond the current window, CascadeLSM maintains a parallel backup LSM-tree that preserves detailed data without aggregation, with application-specific retention.

\subsubsection{Compaction Strategies}
Different tiers employ different compaction strategies optimized for their respective data characteristics and access patterns. For streaming data in the upper levels, we trigger compaction based on capacity thresholds rather than time boundaries. When the memtable reaches a configured size (e.g., 10 GB), it flushes to Level 0 as an SST file. Direct appends through LSM merge operations remove full-sequence read-before-write from foreground ingestion. Within a single processing window, data may compact multiple times between upper levels as capacity thresholds are exceeded.

The key insight of CascadeLSM lies in how we handle the transition from streaming tiers to batch tiers. When a processing window closes, we trigger an aggregation compaction that transforms data from the upper levels to the lower levels. This compaction invokes an external \texttt{CompactionService} that implements a custom \texttt{PartitionKeyMergeOp}. This operator performs user-centric aggregation by treating uid as the partition key during multi-way merge: data is first partitioned by user, then within each partition, records sharing the same \texttt{uid} and \texttt{gid} (item identifier) are aggregated within the processing window. The aggregation logic is application-specific, including deduplication, merging interaction attributes (e.g., taking maximum watch time, bitwise-OR of action flags), and computing aggregate statistics. This design shifts aggregation from the write path, where our previous read-before-write system suffered severe amplification, to the background compaction path, transforming what was an online bottleneck into an offline batch operation.

For batch data in the lower levels, we apply standard capacity-based compaction. When a level reaches its capacity threshold, overlapping SST files are merged to reduce space amplification and maintain read performance. Since batch data is already aggregated, these compactions do not invoke external aggregation logic. Importantly, the aggregated data preserves the read efficiency characteristics of the original KKV (key-key-value) format: a single point lookup by uid retrieves all aggregated interactions for that user within a session window, enabling efficient sequence retrieval for training sample generation without requiring multiple random reads across scattered records.

This cascading design makes compaction both a sorted-run maintenance operation and a reusable sequence transformation. Readers combine the materialized base with recent increments beyond its coverage. Publishing a new base advances that coverage together with the readable file set; the covered staging prefix can then be retired. The processing window chooses when aggregation runs, while event time continues to select relevant behavior. A late event or backfilled observation enters a new processing version even when its event timestamp belongs to an older query interval.

Let $m$ increments accumulate between consolidations, $C(m)$ be consolidation cost, and $R(m)$ the pending work per read. At update/read rates $\lambda_w$/$\lambda_r$, maintenance-dependent work is
\begin{equation}
\mathcal W(m)=\lambda_w C(m)/m+\lambda_r R(m).
\end{equation}
Capacity triggers bound pending state and file count; window triggers select the application transformation boundary. This separates maintenance amortization from history representation.

\begin{tcolorbox}
\vspace{-1mm}
\noindent {\bf Production deployment experience:} In production, compaction serves purposes beyond merging sorted runs. It provides a natural interception point for data transformation such as aggregation and reorganization without additional data movement. By externalizing this logic to a pluggable CompactionService, USDB decouples application-specific transformations from the storage engine core, allowing a single engine to serve heterogeneous upstream and downstream requirements effectively.
\vspace{-1mm}
\end{tcolorbox}

\subsection{Differentiated Replicas}
\label{sec:disagg}

To address the resource imbalance challenge, we build a differentiated multi-replica architecture atop HDFS-based disaggregated storage (Figure~\ref{fig:sep}). Rather than treating all replicas uniformly, we tailor each replica type to its specific workload characteristics, enabling workload-specific resource provisioning and lightweight inter-replica synchronization that avoids redundant compaction.

\subsubsection{Differentiated Node Types}
After disaggregation, we deploy three types of nodes tailored to different workload characteristics. Read-write replicas handle data ingestion and compaction, consuming from Kafka and persisting SST files to HDFS. These nodes require substantial CPU for compaction but minimal local storage since data resides on HDFS. Read-heavy replicas serve inference and streaming-training workloads that require fresh observations. These nodes perform minimal compaction, primarily maintaining a local memtable buffer while reading the majority of data from shared HDFS storage. Read-only replicas serve offline training workloads that can tolerate slightly stale data. These nodes perform no compaction and simply read from HDFS snapshots.

This differentiation allows us to provision each node type with appropriate resources. Read-write nodes receive high CPU allocation with minimal SSD. Read-heavy nodes receive moderate CPU with fast network connectivity. Read-only nodes receive minimal CPU with high memory for caching.

\subsubsection{Replica Consistency}
Read-heavy replicas reuse the writer's persistent state and keep a recent staging region for online workloads. We achieve this through a two-region architecture within each read-heavy replica: a staging region and a shared region.

The staging region subscribes to Kafka and buffers incoming data in a local memtable. This preserves recent observations for fresh reads while shared persistence catches up. When the memtable exceeds a configured size, the replica initiates a flush but first queries the shared region for its maximum Log Sequence Number (LSN). If the shared region's LSN exceeds the staging region's maximum LSN, the memtable can be safely discarded since the data is already persisted. Otherwise, only the delta beyond the shared region's LSN needs to be flushed to HDFS. Read-heavy replicas may therefore perform fallback persistence and maintenance when the shared writer lags.

The shared region opens the HDFS-resident RocksDB instance in read-only mode using the OpenAsFollower mechanism. It periodically applies metadata changes from the read-write replica, tracking the maximum persisted LSN. During read operations, a MergeRead with LSN filtering combines data from both regions, using the shared region's LSN as a filter to exclude redundant entries from the staging region.

This design ensures that read-heavy replicas observe a consistent data state while minimizing redundant compaction work. The staging region supplies recent observations, while the shared region supplies the persisted prefix. For one ingestion partition, a shared frontier $p$ and staging frontier $q$ give the read view as shared records through $p$ plus staging records in $(p,q]$. Advancing $p$ with the readable file set permits covered staging data to be reclaimed.

\subsubsection{Simplified Scaling}
Disaggregation dramatically simplifies cluster scaling. In the previous architecture, scaling required copying complete shard data to new nodes. With disaggregation, we construct a separate database for each Kafka partition within a shard. Scaling then proceeds as follows: first, new read-only nodes load the target partition databases from HDFS; second, the old read-write nodes unload those partitions; third, new read-write nodes load the partitions and begin consuming from Kafka; finally, old read-only nodes unload the partitions. Historical SST files remain on HDFS and can be opened by the new owners; the handoff reuses the existing bulk data.

Opening the target database, replaying its uncovered log suffix, and warming the reader cache determine readiness after handoff. Bulk history files stay reusable in shared storage.

\begin{tcolorbox}
\vspace{-1mm}
\noindent {\bf Production deployment experience:} While disaggregated storage is increasingly common in modern KV systems~\cite{dong2023disaggregating,shan2018legoos}, the key challenge lies in how replicas are organized and kept consistent atop shared storage. By differentiating nodes into replicas with workload-specific provisioning, and coordinating them through lightweight synchronization, USDB avoids redundant compaction across replicas while maintaining consistency.
\vspace{-1mm}
\end{tcolorbox}

\begin{figure}[t]
    \centering
    \includegraphics[width=\linewidth]{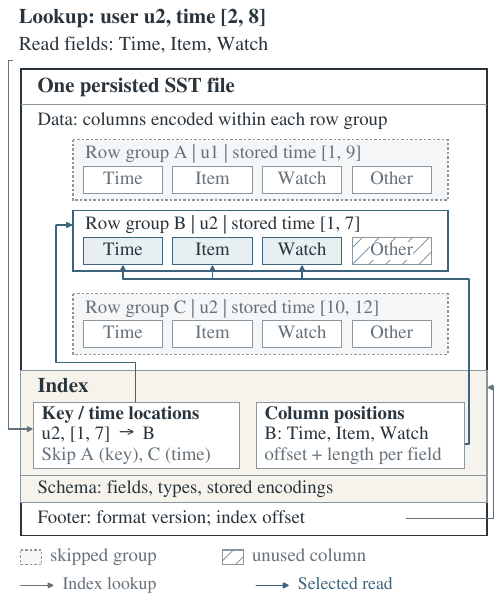}
    \caption{Indexed columnar layout: data blocks, column metadata, and per-user access share one file format.}
    \label{fig:format}
\end{figure}

\subsection{Partial Dump with MVCC}
\label{sec:mvcc}

To achieve high training-inference consistency while controlling storage costs, we design a two-part solution that balances these competing requirements. As discussed in Section~\ref{sec:inconsist}, conventional full sample dump approach becomes prohibitively expensive at our scale. We address this through partial dump for cost efficiency combined with Multi-Version Concurrency Control (MVCC) based on Log Sequence Numbers (LSNs) to select the observations available to inference.

\subsubsection{Partial Dumps}
Our partial dump mechanism captures only the most recent portion of user behavior sequences during inference, rather than dumping complete sequences. Complete training samples are then reconstructed by merging partial dumps with historical data stored in the training cluster. However, partial dump alone reintroduces consistency challenges that full dumps inherently avoided. A naive implementation uses the user's event timestamp (when the action occurred) to align partial dump data with historical data during the merge. This works correctly when out-of-order events appear only at sequence boundaries, as shown in Figure~\ref{fig:dump}(a). The problem arises when out-of-order events fall within the middle of a sequence. Consider Figure~\ref{fig:dump}(b) where actions 4 and 5 arrive out of order due to network lag. During inference, the online sequence observes [2,3,6,7,8] while the training cluster has received [0,1,2,3,4,5,6,7,8]. The partial dump captures [6,7,8] with inclusive timestamp range [2,8]. During training, merging based on timestamps produces [2,3,4,5,6,7,8], which includes actions 4 and 5 that were not visible during inference. This training-inference inconsistency degrades model quality.

\subsubsection{Versioned Replay}
We resolve this through MVCC by assigning each action an LSN representing its system arrival time rather than its event time. As shown in Figure~\ref{fig:dump}(c), each action carries both its event timestamp and its LSN. The online sequence (2,3,6,7,8) has LSNs (c,d,e,f,g). The training sequence (0,1,2,3,4,5,6,7,8) has LSNs (a,b,c,d,h,i,e,f,g), where actions 4 and 5 received later LSNs (h,i) despite having earlier event timestamps. During partial dump, we record both the timestamp range and the LSN range. During training, we filter the training sequence to include only actions whose LSNs fall within the dumped LSN range. Actions 4 and 5 have LSNs (h,i) that exceed the maximum dumped LSN (g), so they are correctly excluded. The resulting training sequence (2,3,6,7,8) matches the inference sequence exactly. The saved version boundary likewise excludes observations introduced by a later retransmission. The source partition and version domain travel with the reconstruction metadata so that the cutoff is applied to the corresponding history.

\paragraph{Retaining the right version.}
For an aggregate whose values evolve, reconstruction also selects a compatible retained base or original detail. For example, watch-time observations 2 at $v_1$ and 5 at $v_2$ produce a latest maximum of 5; replay bounded by $v_1$ uses the earlier observation 2. Retaining detailed inputs or version-compatible materializations supplies that earlier value. Event time chooses behavior, ingestion version chooses available observations, and the published file view identifies the readable representation. Imported feature versions identify their corresponding historical base.

The pipeline records version ranges and action counts and exposes reconstruction consistency to downstream consumers. Applications choose a partial window that bridges recent pipeline delays and their recovery interval. A larger captured window stores more of the inference state directly; a smaller one increases historical reuse. The same interface supports this choice without changing the training sample's logical input.

\begin{tcolorbox}
\vspace{-1mm}
\noindent {\bf Production deployment experience:} While recommendation, advertising, and search ranking systems tolerate weak consistency in the traditional database sense, high training-inference consistency remains critical for model accuracy. Full sample dumps ensure consistency but incur prohibitive storage overhead at scale. USDB pairs lightweight partial dumps with LSN-based versioning for high consistency and lower snapshot cost.
\vspace{-1mm}
\end{tcolorbox}

\subsection{Hybrid Columnar Encoding}
\label{sec:encoding}
Typical LSM implementations use row-oriented SST formats optimized
for point lookups, but this organization performs poorly for training
workloads that scan specific feature columns across millions of user
sequences. To address this, we developed a hybrid columnar format that
balances inference and training access patterns, as illustrated in
Figure~\ref{fig:format}.

\subsubsection{Self-Describing Format}
Our file format comprises three regions: a data region, an index
region, and a footer. Inspired by PAX~\cite{ailamaki2001weaving} and
Parquet, the data region partitions data into row groups
and stores values column-by-column within each group, but further
introduces a fine-grained layout hierarchy for flexible IO and
predicate pushdown. Concretely, the data region is organized as a
\texttt{RootLayout} containing one \texttt{ZoneLayout} per column, each
divided into \texttt{ChunkLayout}s (analogous to column chunks within a
row group). Each \texttt{ChunkLayout} groups multiple
\texttt{FlatLayout(data)} blocks, the basic unit of IO and computation.
Each \texttt{ZoneLayout} also maintains a dedicated
\texttt{FlatLayout(zone)} storing per-column zone maps that enable
predicate pushdown and fine-grained IO skipping. The index region
includes \texttt{Zone Maps \& Range Indexes} for range queries,
\texttt{BloomFilter} for point lookup acceleration, and
\texttt{TableScheme} for self-describing metadata. The footer stores
\texttt{FileVersion} and \texttt{IndexOffset} for format versioning and
index location. Embedded schema and file-version metadata identify stored fields and support additive column evolution. Applications supply feature identity and aggregation semantics. Column-oriented decoding feeds vectorized data processing while preserving the same indexed files for serving.

Aggregation pushdown executes an aggregate over the selected user history near the stored data and returns a compact result instead of transferring feature rows. Our implementation supports approximate count while retaining the query's user, time-range, and Top-$K$ selection. Section~\ref{sec:count-pushdown} evaluates this aggregate under increasing query concurrency and constrained receive bandwidth.

\subsubsection{Adaptive Encoding Selection} Our format employs a
decision tree-based encoding selector that chooses
encodings based on column types and data characteristics. Inspired by
recent columnar compression works~\cite{kuschewski2023btrblocks,
vortex2024}, the selector cascades lightweight encodings, including
dictionary encoding, frame-of-reference, bit-packing, and run-length
encoding, adaptively based on column statistics such as cardinality,
value distribution, and data type. As shown in Section~\ref{sec:compress},
this approach achieves significantly faster decompression than
general-purpose compressors like ZSTD and LZ4 while maintaining
encoded sizes between the tested LZ4 and ZSTD configurations. This trade-off is critical for training
workloads where decompression throughput directly impacts data loading
performance.

The hybrid columnar encoding serves both access patterns: columnar scans with lightweight encoding
for training throughput, and row-level indexing for inference latency.
\begin{tcolorbox}
\vspace{-1mm}
\noindent {\bf Production deployment experience:} File format flexibility is essential for serving diverse workloads
within a unified storage system: smaller row groups favor point lookups
for online inference, while larger row groups improve compression and
scan throughput for offline training. USDB's self-describing format
enables the storage engine to adapt layout and encoding decisions to
workload characteristics without requiring separate storage paths.
\vspace{-1mm}
\end{tcolorbox}

\FloatBarrier
\setcounter{topnumber}{2}
\setcounter{dbltopnumber}{2}
\setcounter{totalnumber}{3}
\section{Production Evaluation}
\label{sec:eval}
We evaluate four design questions: maintenance efficiency, replica provisioning, repeated sample payload, and selective history reads (Table~\ref{tab:evidence}). We first define the query patterns and compare the deployed read paths under controlled load, then evaluate resource accounting, reconstruction, and encoding. Production scale and full-day operation close the empirical study.

\begin{table}[t]
\footnotesize\centering
\caption{Challenge $\rightarrow$ solution $\rightarrow$ evidence.}
\label{tab:evidence}
\setlength{\tabcolsep}{2pt}
\renewcommand{\arraystretch}{1.16}
\begin{tabular}{@{}lll@{}}
\toprule
Challenge & Solution & Evidence \\\midrule
Growing histories & Time-based stages & Write/CPU traces \\\addlinespace[2pt]
Replica overhead & Shared files & Provisioning/cost \\\addlinespace[2pt]
Repeated dumps & Tail + versioned prefix & Reconstruction \\\addlinespace[2pt]
Large reads & Columnar + pushdown & Read/aggregate QPS \\
\bottomrule
\end{tabular}
\vspace{-2mm}
\end{table}

\subsection{Read Performance}
\label{sec:controlled}
\textbf{Systems and workload.} We compare USDB with \emph{Optimized RocksDB}, a deployed sequence service with production indexing and column selection. Both systems run on matched hardware with the same deployment topology and data-center environment. Both use comparable production-workload samples, 15 common fixed-width columns, and identical query parameters.

\textbf{Method.} Queries retrieve a user's history within an event-time range, in descending time order, with a Top-$K$ limit and selected columns. Both clients use the same configuration and UID-sampling rule. Ordinary runs use 30 seconds of warm-up followed by up to 60 seconds of measurement; extended sweeps also cap each phase at 400K requests. Fixed-work jobs time retrieval under equal request budgets, including startup and shutdown, before feature decoding and model computation.

The study contains 374 measurement runs across history length, projection, load, client resources, and access locality. Throughput is the median QPS over active five-second reporting intervals; latency uses cumulative request statistics. Errors and empty responses are included in the outcome analysis. Repeated conditions report mean $\pm$ sample standard deviation over three independent runs ($n=3$), with individual points shown.

\begin{figure*}[t]
\centering\includegraphics[width=\textwidth]{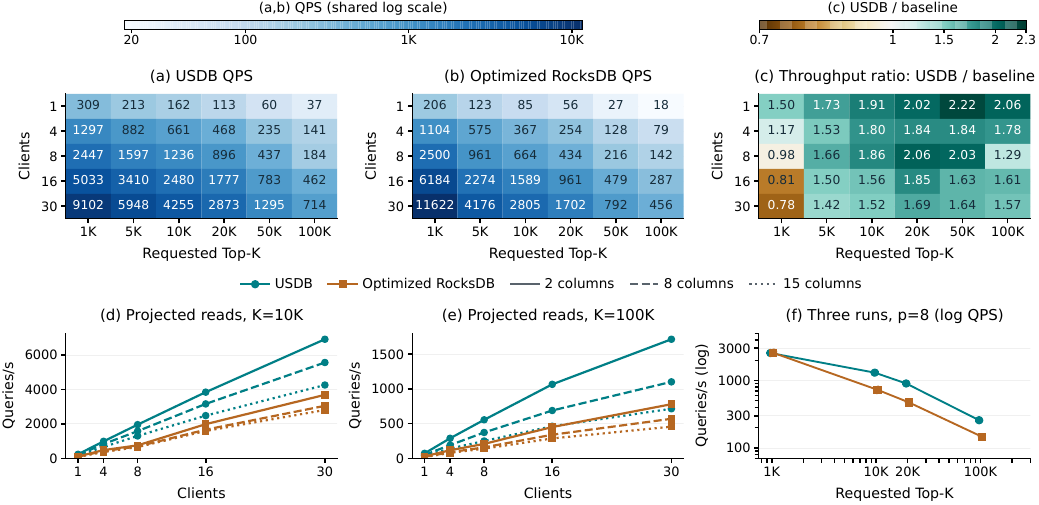}
\caption{History length, projection, and client load. (a--b) QPS for 15 columns, with a shared logarithmic color scale. (c) USDB/baseline throughput ratio; 1 denotes parity. Color keys appear above the heatmaps; line keys apply to the lower row. (d--e) Two/eight/15-column reads. (f) Three runs per length at eight clients: mean $\pm$ sample SD, with individual points and logarithmic QPS.}
\label{fig:controlled}
\end{figure*}

\subsubsection{Length and Load}
USDB's advantage grows as reads retrieve more history (Figure~\ref{fig:controlled}). At eight clients, throughput is $1307.0\pm15.7$ versus $728.0\pm23.6$ QPS for 10K-action requests (1.79$\times$), and $257.7\pm2.6$ versus $146.8\pm0.3$ for 100K (1.75$\times$), using three runs per system and length ($n=3$). At 20K, throughput is $903.7\pm6.1$ versus $473.3\pm3.1$ QPS (1.91$\times$). The 100K mean latency is 31.13 versus 54.47 ms; both systems return roughly 90K rows per nonempty result. The same repetitions give similar 1K throughput, $2554.3\pm21.5$ versus $2578.0\pm79.5$ QPS. This progression connects the benefit to repeated long-history retrieval.

The crossed scan extends the result across five client loads and six requested lengths. At 30 clients, USDB reaches 4255/714.5 QPS at 10K/100K, compared with 2805/456.5: 1.52$\times$/1.57$\times$. At one client, the corresponding ratios are 1.91$\times$/2.06$\times$. Short 1K requests scale differently: at 30 clients the baseline reaches 11622.5 QPS and USDB 9102.5, with both reaching the request budget. An independent 100K concurrency sweep reaches 654.5/441.5 QPS at 30 clients and p99 100.67/102.89 ms. 

\subsubsection{Projection across Lengths}
Selecting fewer columns increases useful long-read capacity in both systems. At 100K and eight clients, one versus 15 columns increases USDB throughput from 250.5 to 598 QPS and the baseline from 139.5 to 210.5; p99 falls from 51.11 to 26.08 ms and from 91.20 to 63.51 ms. At 30 clients, two/eight/15 columns yield USDB 1715/1102/714.5 QPS versus 780/570/456.5, giving 2.20$\times$/1.93$\times$/1.57$\times$ (Figure~\ref{fig:controlled}e).

\begin{figure*}[t]
\centering\includegraphics[width=\textwidth]{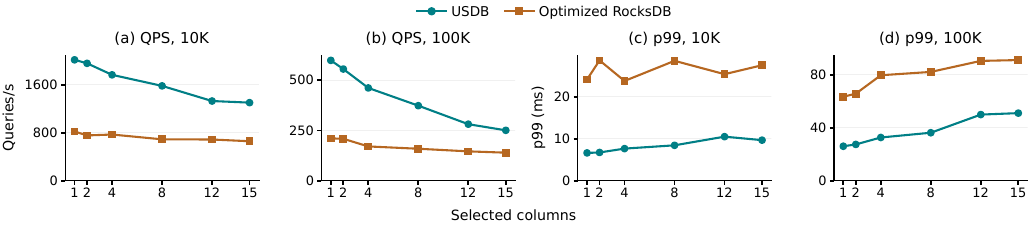}
\caption{Projection at eight clients across six column selections. (a--b) Throughput; (c--d) p99 latency. Each metric is shown separately for 10K- and 100K-action reads.}
\label{fig:projection}
\end{figure*}

The six-width scan separates output width from requested length (Figure~\ref{fig:projection}). Long reads benefit from reducing the materialized payload; short prefixes contain a larger share of fixed lookup and request work. This explains why one-column selection is not a simple inverse-width speedup. Three independent runs at 10K/100K and two/eight columns further check projected-query variation. Together, the crossed load/width curves and repetitions show where selective materialization contributes within each complete read path.

\subsubsection{Client Scaling}
\begin{figure*}[t]
\centering\includegraphics[width=\textwidth]{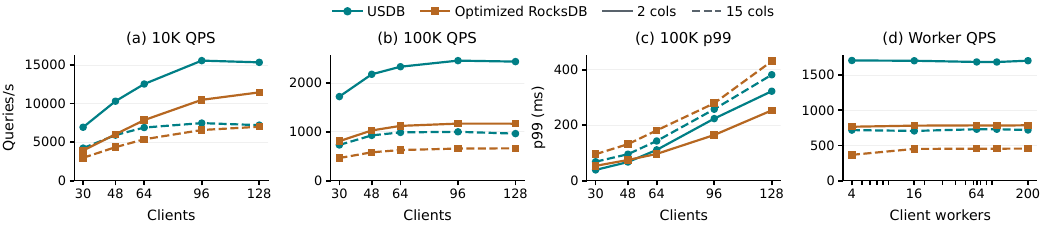}
\caption{Client scaling with two/15 columns. (a--c) Throughput and p99 across client loads. (d) Client-worker sensitivity at 100K and 30 clients.}
\label{fig:clientcapacity}
\end{figure*}

Figure~\ref{fig:clientcapacity} increases concurrency from 30 to 128 clients for 10K/100K reads and two/15 columns. At 100K and 96 clients, USDB reaches 2457.5 QPS with two columns and 1001 with 15, compared with 1169 and 658.5. Beyond 96, throughput largely plateaus while p99 rises. At 128, two-column p99 is 322.62 versus 253.48 ms, and 15-column p99 is 381.88 versus 430.70 ms. The curves distinguish throughput capacity from the latency operating point of a consumer.

We also vary client workers from 4 to 200 and read buffers from 4 to 256 KiB. For 15-column 100K reads, the baseline reaches 368/455.5/456.5 QPS with 4/100/200 workers; USDB reaches 719.5/731.5/721. The sweep reaches a stable throughput region, indicating sufficient client parallelism. Buffer changes produce less than 2\% throughput range for each system/width. USDB's wide-query client CPU cost is 3.95/5.69 ms per completed attempt at 4/100 workers, showing why extra workers need not improve useful capacity.

Five repeated default settings per system and projection quantify variation across the client sweeps. Mean 100K, 30-client throughput is 1709.4/789.5 QPS for two columns and 725.8/459.6 for 15. The client study records six/45 request errors for USDB/baseline; USDB also records 316 empty results. The companion outcome analysis reports the per-load error and empty-response rates.

\subsubsection{Read Jobs and Mixed Loads}
The fixed-work jobs test completion of a known amount of input reading. At eight clients, the longer jobs issue exactly 150K, 50K, or 30K requests for requested lengths 1K, 10K, or 100K. Each request succeeds with a nonempty result. At 100K, two/four/eight/15-column completion times are 55.82/65.85/80.78/120.81 seconds for USDB and 130.74/166.01/170.64/195.69 for the baseline: 2.34$\times$/2.52$\times$/2.11$\times$/1.62$\times$ speedups. At 10K, the corresponding speedups are 1.82$\times$/2.13$\times$/1.70$\times$/1.61$\times$.

\begin{figure*}[t]
\centering\includegraphics[width=\textwidth]{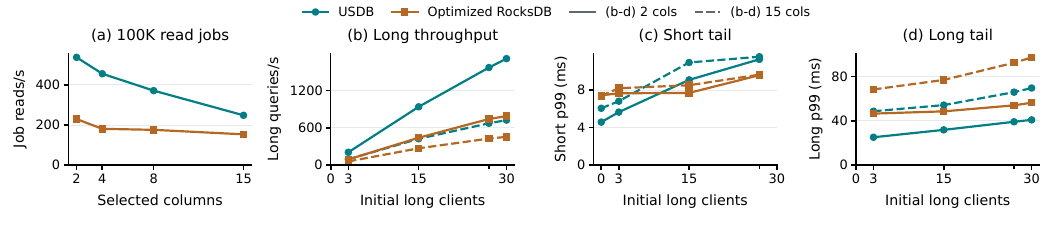}
\caption{Read jobs and mixed loads. (a) 30K-request jobs on 100K-action histories, eight clients. (b--d) Long-read QPS and short/long p99 for co-started 1K/100K reads sharing 30 clients; each stream runs for up to 60 seconds or 400K requests. Solid/dashed lines indicate two/15 columns.}
\label{fig:readjobs}
\end{figure*}

Figure~\ref{fig:readjobs}a shows the longer 100K-history jobs. At the shorter 10K-request budget, the corresponding jobs take 21.01/30.87/40.89 seconds versus 50.87/66.01/71.17. Longer jobs amortize fixed startup and reporting exit costs, preserving the long-read advantage. At 1K and eight/15 columns, longer-job times are 51.61/60.83 versus 50.82/56.03 seconds, consistent with the different balance of fixed and payload work for short reads.

Mixed consumers co-start short (1K) and long (100K) processes with initial client allocations 30:0, 27:3, 15:15, 3:27, and 0:30. Each stream has its own worker pool and bounded budget. Both streams execute concurrently; short streams can finish earlier when they reach the request budget. At 15:15 with 15 columns, USDB's long stream reaches 423.5 QPS and p99 54.08 ms, versus 270 and 76.78 ms. Its short stream reaches 3876 QPS and p99 10.94 ms, versus 4436 and 8.51 ms. The combined view shows how long-read gains coexist with the latency needs of a fresh, short-history consumer.

\subsubsection{Sustained Reads and Locality}
\begin{figure*}[t]
\centering\includegraphics[width=\textwidth]{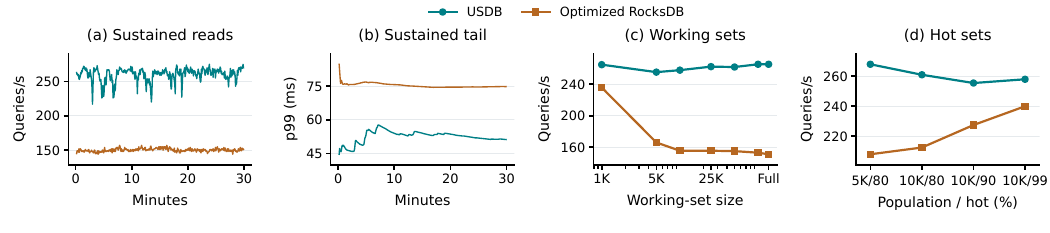}
\caption{Read stability and locality, eight clients, 100K, 15 columns. (a--b) Throughput and cumulative p99 from a 30-minute run; the initial report and incomplete final interval are omitted from the traces. (c) Uniform working-set throughput. (d) Throughput versus population/hot-request share; the hot set contains 1K users. Time-window results appear in Table~\ref{tab:windows}.}
\label{fig:sustained}
\end{figure*}

The 30-minute runs at eight clients, 100K, and 15 columns complete with zero errors or empty results. Median QPS is 263/150; cumulative mean latency is 30.69/53.09 ms and p99 51.24/74.81 ms. Figure~\ref{fig:sustained} places the sustained report series beside the population and hot-set scans.

Increasing the working set from larger uniform samples to the full sampled workload yields USDB 262--265.5 QPS and baseline 151--155.5. Keeping the same 10K subset while increasing the 1K hot-set share from 80\% to 99\% raises baseline throughput from 212.5 to 240; USDB changes from 261 to 258. The scan also includes smaller working sets. These distributions vary reuse while keeping the requested output fixed.

\begin{table}[t]
\small\centering
\caption{Lookback windows, paired USDB/Optimized RocksDB values. Rows/hit is rounded down. $^*$Empty-window baseline latency uses the final reporting window.}
\label{tab:windows}
\setlength{\tabcolsep}{2.5pt}
\begin{tabular}{@{}lrrrr@{}}
\toprule
Window & Empty (\%) & Rows/hit & Mean ms & p99 ms \\
\midrule
1h & 100.00/100.00 & --/-- & 2.33/0.41$^{*}$ & 3.77/0.57$^{*}$ \\
6h & 100.00/100.00 & --/-- & 2.36/0.56$^{*}$ & 3.81/2.70$^{*}$ \\
1d & 34.22/29.95 & 90/90 & 2.44/1.74 & 3.85/8.73 \\
3d & 0.88/0.86 & 840/1007 & 2.92/3.31 & 4.90/11.79 \\
7d & 0.21/0.35 & 2359/2819 & 3.48/5.00 & 6.54/15.01 \\
\bottomrule
\end{tabular}
\end{table}

Time-window selection changes the amount of eligible history (Table~\ref{tab:windows}). One-hour and six-hour windows are empty at the fixed endpoint. One-day nonempty results average about 90 rows, and seven-day results 2359/2819 despite the 100K maximum. The baseline has lower mean latency for short windows; USDB has lower means at three/seven days and lower one/seven-day p99. Reporting coverage and actual rows explains these points without treating a requested maximum as the returned work.

\subsection{Aggregation Pushdown}
\label{sec:count-pushdown}
We evaluate aggregation pushdown using approximate count over selected histories. USDB returns the aggregate as a scalar; the two comparison paths count timestamp rows retrieved through ordinary USDB and Optimized RocksDB reads. All paths use matched histories, the same time window, and the same Top-$K$ cap. Sampled correctness checks agree exactly in 77 of 80 cases, with the remaining estimates differing by at most two rows.

\begin{figure}[t]
\centering\includegraphics[width=\columnwidth]{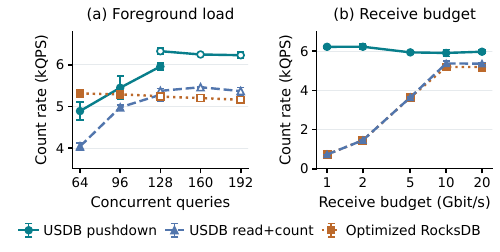}
\caption{Aggregation throughput with all client CPUs available; count queries. (a) Foreground QPS; filled/open markers: 128/256 workers. (b) QPS versus receive budget, 128 clients/workers. Means $\pm$ sample SD ($n=3$).}
\label{fig:count-load}
\end{figure}

\textbf{Matched histories under load.} We fix a length-matched user cohort and a 30-day window, returning about 33K actions per query with a 10M Top-$K$ cap. All client CPUs remain available. The foreground sweep uses eight seconds of warm-up and 20 seconds of measurement, repeated three times; QPS is the per-run median of active two-second reports. Figure~\ref{fig:count-load}a uses fixed 128 workers at 64/96/128 concurrent queries and fixed 256 workers at 128/160/192 queries, with both sweeps including the 128-query setting. At 192 queries, native count reaches 6.23K QPS versus 5.37K for USDB read-and-count and 5.16K for Optimized RocksDB, improving throughput by 1.16$\times$/1.21$\times$.

\textbf{Bandwidth-constrained counting.} We fix 128 concurrent queries and 128 workers, then vary the aggregate client receive budget from 1 to 20 Gbit/s across the same three paths (Figure~\ref{fig:count-load}b). All client CPUs remain available, and socket counters independently verify the receive rate. Each point has 15 seconds of warm-up and 45 seconds of measurement, repeated three times. Successful QPS is the median active report rate after excluding failed and empty responses. At 1 Gbit/s, pushdown sustains $6.22$K QPS versus about $0.73$K for either scan path, yielding 8.52$\times$/8.54$\times$ higher throughput. At 5 Gbit/s, the gains are 1.63$\times$/1.63$\times$. The scalar response requires only about 10 Mbit/s at 1 Gbit/s budget; ordinary reads exhaust the smaller budgets and approach unrestricted throughput as the budget increases. Pushdown preserves throughput when result transfer limits ordinary reads.

\subsection{Replica Provisioning and Cost}
\label{sec:dis_cost}
Differentiated replicas change the provisioning unit from a complete private storage-and-maintenance stack to a workload role. Table~\ref{tab:cost} reports deployment accounting for local disk on memory-oriented machines and shared storage on balanced machines. At a normalized 1M reads/s and 50\% CPU utilization, compute cost falls from 2.73 to 1.00, storage from 1.93 to 1.70, and total from 4.66 to 2.70 (42.1\%). CPU provisioning falls from 6284 to 4562 cores and memory from 62840 to 9124 GiB.

\begin{table}[t]
\small\centering
\caption{Deployment provisioning and normalized cost. Shared-storage compute is the unit; the compared machine classes reflect their workload roles.}
\label{tab:cost}
\begin{tabular}{@{}lrr@{}}
\toprule
Metric & Local disk & Shared storage \\\midrule
CPU cores & 6284 & 4562 \\
Memory (GiB) & 62840 & 9124 \\
Core:memory ratio & 1:10 & 1:2 \\
Throughput (K/s/core) & 0.32 & 0.44 \\
Compute cost & 2.73 & 1.00 \\
Storage cost & 1.93 & 1.70 \\
Total cost & 4.66 & 2.70 \\\bottomrule
\end{tabular}
\vspace{-2mm}
\end{table}

The larger reduction in compute and memory reflects removing repeated maintenance from reader provisioning and selecting hardware by role. Persistent copies, writer compute, reader caches, and retention remain explicit cost components. The organization also lets an additional reader reuse existing historical files rather than first create a complete private shard image. This connects the publication protocol in Section~\ref{sec:disagg} to an operational benefit beyond request throughput.

\subsection{Reconstruction Consistency}
\label{sec:partial_analysis}
\begin{figure}[t]
\centering\includegraphics[width=\columnwidth]{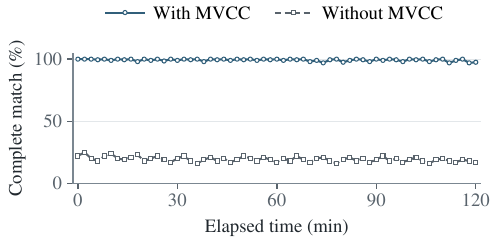}
\caption{Versioned reconstruction. Complete-sequence agreement over two hours, with and without MVCC.}
\label{fig:mvcc}
\end{figure}

The application study compares complete-sequence agreement with and without arrival-version filtering over two hours (Figure~\ref{fig:mvcc}). All actions in a reconstructed sequence must agree with the inference input for a match. Across the 61 reported observations, matching is 97--100\% with MVCC and 16--25\% without it; 27 MVCC observations reach 100\%. The later historical view contains observations that were unavailable to inference, and the saved arrival-version boundary selects the relevant state. Partial dumps capture recent state directly, while the versioned history supplies its reusable prefix.

One production policy captures the most recent day, covering the dominant late-arrival interval and the typical outage-detection/recovery window. The deployment reports a captured sample fraction of 10--20\%, motivating reuse of the older history during reconstruction. The following accounting includes the captured fragment and its metadata explicitly.

For $N$ samples, let $F$ denote average encoded bytes of a full sequence, $P$ the captured sequence bytes, $M$ reconstruction metadata, and $O$ unchanged sample features. Sample storage becomes
\[
B_{\rm full}=N(O+F),\qquad B_{\rm partial}=N(O+P+M).
\]
The sequence-payload reduction is $1-(P+M)/F$, and entire-sample reduction is $(F-P-M)/(O+F)$. This accounting exposes the benefit directly: overlapping historical prefixes are retained once in shared history rather than copied into every sample. Overall deployment accounting additionally includes the retained versions and detailed history used for reconstruction.

Using mean captured/original lengths 195.7/2950, the storage model yields a 6.6\% captured-length ratio. A separate 24-hour pipeline study of 2K sequences compares nine dump-window policies. One policy captures at most 660 actions under 30-second averaging, with item--timestamp agreement of 0.999--1.000; both metrics report the worst window of the day. These results guide the captured window toward recent pipeline delays while reusing the larger historical prefix. Length quantifies capture volume; item--timestamp agreement measures pipeline alignment; complete-sequence agreement measures reconstruction correctness.

The controlled read jobs quantify the retrieval side of this design. Their 100K completion improvement persists across narrow and wide projections, showing how lower lookup time can support a sample pipeline that materializes history later. The storage model and read experiments thus address complementary parts of the same lifecycle: reduce repeated bytes at sample creation and supply reusable history efficiently at consumption.

\subsection{Compression Performance}
\label{sec:compress}
\begin{figure}[t]
\centering\includegraphics[width=\columnwidth]{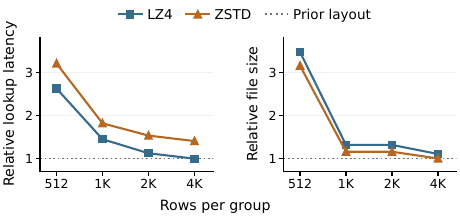}
\caption{Row-group size versus lookup latency and file size, normalized to the prior layout (dotted reference). K denotes thousands of rows per group.}
\label{fig:codecstudy}
\end{figure}

The physical-layout study uses a sample of production histories, keeping key columns separate and grouping value columns. Its reference layout takes 9994.45~$\mu$s and 1.9G. Figure~\ref{fig:codecstudy} compares nine configurations: LZ4/ZSTD with 512, 1,024, 2,048, or 4,096 rows per group, plus the reference layout.

At 4K rows, LZ4 takes 9957.65~$\mu$s and 2.1G, approaching reference latency with 10.5\% more bytes; ZSTD uses 1.9G at 14045.65~$\mu$s. At 512 rows, LZ4 takes 26269.50~$\mu$s and 6.6G. Finer groups increase repeated metadata and fragmented reading in this study. The complete size/latency curve explains why grouping must be selected with the user-history access pattern, complementing the controlled projection scan.

The codec benchmark (Table~\ref{tab:compression}) shows the materialize-once/read-repeatedly choice. USDB decodes 4.6--5.5$\times$ faster than ZSTD, using 27.9--42.2\% more encoded bytes and lower encode throughput. At 256K groups, it decodes at 14.88 GB/s versus 6.88 for LZ4 and 3.21 for ZSTD. Background construction amortizes encoding over repeated queries; the projection and read-job results evaluate the corresponding application-facing path.

\begin{table}[t]
\footnotesize\centering
\caption{Codec measurements. Ratio is encoded/original size; throughput is measured at the codec boundary.}
\label{tab:compression}
\setlength{\tabcolsep}{3pt}
\begin{tabular}{llrrr}
\toprule
Group & Method & Ratio & Encode MB/s & Decode GB/s \\\midrule
1K & USDB & 12.8\% & 154.5 & 9.05 \\
 & LZ4 & 13.8\% & 1950.1 & 5.36 \\
 & ZSTD & 9.9\% & 637.6 & 1.66 \\
8K & USDB & 12.8\% & 407.1 & 13.80 \\
 & LZ4 & 13.1\% & 2072.7 & 5.38 \\
 & ZSTD & 9.0\% & 1083.8 & 2.57 \\
256K & USDB & 11.0\% & 420.4 & 14.88 \\
 & LZ4 & 12.6\% & 2403.4 & 6.88 \\
 & ZSTD & 8.6\% & 1297.9 & 3.21 \\
\bottomrule
\end{tabular}
\vspace{-2mm}
\end{table}

\subsection{Direct-File Training Access}
The production training study compares service-mediated access and direct shared-file reads for the same model, with 2,000 data-reader workers and 16 training nodes in both configurations and caching enabled. Trainer-reported throughput increases from 175K to 184K instances/s (5.1\%), while weighted training-job CPU utilization remains about 32\%. Cache hit rate increases from 92\% to 96\%; mean uncached sequence-fetch latency is 3.3/3.4 ms. Direct-file access reuses the materialized historical representation in the training pipeline. These training results complement the controlled fixed-work reads in Section~\ref{sec:controlled}, which measure client reading separately from model computation.

\subsection{Production at Scale}
\label{sec:production}
A September 2026 deployment record lists approximately 18 PB of retained history per region, including two copies, across three regions for one application family. This operational footprint complements the representative demand stages in Figure~\ref{fig:traffic}. The migration study below uses full-day records from an earlier deployment.

The production migration replaced read-before-write sequence updates in separate local-disk streaming/batch stores. Its legacy deployment had 256 nodes with 32 cores and 100-Gbps interfaces, retaining about 20 TB of streaming and 600 TB of batch data including replication. This earlier migration baseline is a deployed Lambda-style sequence service with separate streaming/batch databases, user indexing, and aggregation. Optimized RocksDB in the controlled study is the subsequent optimized sequence service and supports projection; the new comparison uses both systems' deployed read paths.

\begin{figure*}[t]
\centering\includegraphics[width=\textwidth]{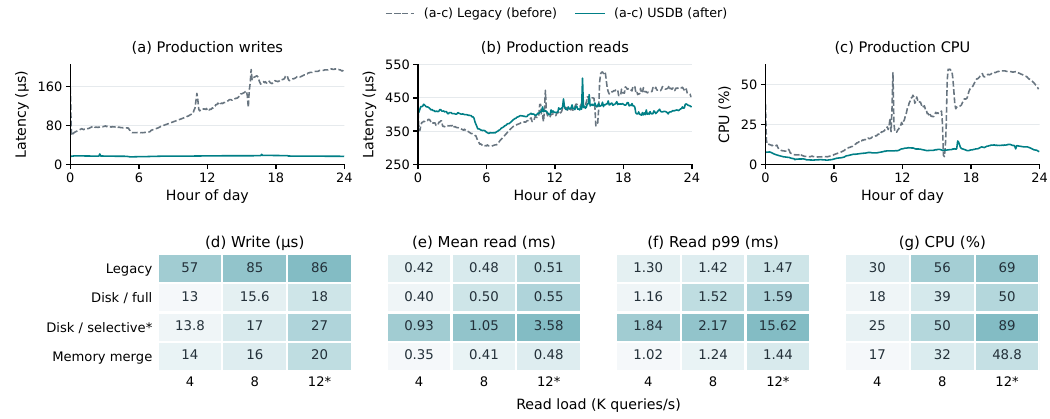}
\caption{Production operation and replica policies. (a--c) Full-day before/after records in five-minute averages. (d--g) Policy/load measurements; darker cells rank higher cost within each metric. Disk rows use full/selective caching; memory uses merge-on-read. *Selective caching uses 11.5K reads/s in the last column; other rows use 12K.}
\label{fig:online}
\end{figure*}

\textbf{Foreground efficiency.} Across the complete 24-hour records for each deployment, mean write latency falls from 123.94 to 17.56~$\mu$s (85.8\%), and mean CPU utilization from 29.04\% to 7.95\% (72.6\%). Mean read latency remains near 410~$\mu$s, changing from 409.71 to 406.15; CPU peaks change from 64.3\% to 14.7\%. Figure~\ref{fig:online} shows the complete daily time series. Foreground append decouples each update from full-history rewriting; background materialization preserves user-local read access. A write contributes one action, whereas an inference read typically retrieves about 1K actions in this workload. Read requests account for roughly 10--20\% of write volume, making the write path a primary source of resource pressure.

\textbf{Maintenance and cache choices.} The single-shard experiment uses a 32-core, 256-GB machine and random reads averaging 1.02K actions. At 12K reads/s, memory merge-on-read uses 48.8\% CPU versus legacy's 69\%, with write latency 20 versus 86~$\mu$s and read p99 1.437 versus 1.474 ms. The configuration study makes pending state and caching visible: the disk path with selective caching reaches 89\% CPU and 15.617-ms p99 at 11.5K reads/s. Dedicated materialization and read-cache provisioning determine the operating point together. Engine write amplification is 55 for legacy and 66 for the USDB configurations; foreground append latency captures the benefit of moving application-level sequence rewriting into background maintenance.

\begin{figure}[t]
\centering\includegraphics[width=\columnwidth]{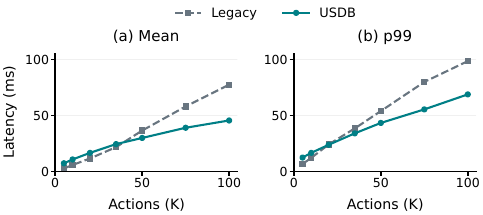}
\caption{Migration-era long reads. Mean and p99 latency across seven sequence lengths.}
\label{fig:legacyreads}
\end{figure}

\textbf{Long histories.} The production long-read study also shows the shift in the cost balance (Figure~\ref{fig:legacyreads}). At 5K actions, mean latency is 7.28/2.71 ms for USDB/legacy. At 100K, it is 45.5/77.5 ms and p99 is 68.8/98.5 ms. The controlled comparison in Section~\ref{sec:controlled} extends this evidence with a stronger deployed baseline, common columns, repetitions, and crossed workload parameters. Both studies make retained history length central to interpreting read performance.

\begin{figure}[t]
\centering\includegraphics[width=\columnwidth]{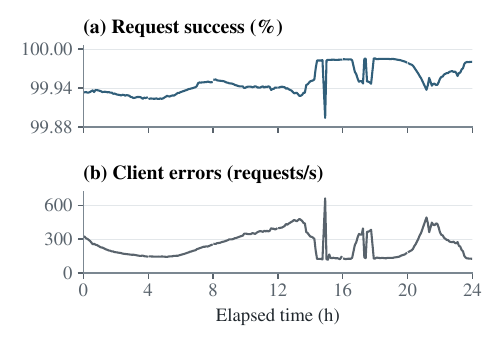}
\caption{Production request outcomes over 24 hours. Gaps mark missing observations.}
\label{fig:production-reliability}
\end{figure}

\paragraph{Request outcomes in production.}
A separate 24-hour production window shows request success of 99.894--99.986\% and client-error volume of 116--661 requests/s (Figure~\ref{fig:production-reliability}). The trace contains 282 paired five-minute observations; six missing intervals appear as gaps. Together with the latency and resource curves, these outcomes characterize the deployment's operating point.

\subsection{Deployment Lessons}
\label{sec:lessons}
\textbf{Share the materialization investment.} Aggregation and encoding both spend work during background construction to reduce subsequent reading. Their operating points should reflect repeated consumption and the read/write mix, alongside retained size. The production traces, codec study, and long-read jobs show these choices at their respective boundaries.

\textbf{Provision for useful work.} Length and width change payload, UID reuse changes caching benefit, and concurrency changes throughput and tail latency. The client study finds little 100K throughput benefit beyond 96 clients; additional workers can increase CPU without increasing reads. Matching these needs to a replica role makes shared capacity useful to each consumer.

\textbf{Make history part of the sample contract.} Recording selection and visibility together lets training reuse history while preserving the recent inference observation. Event time, ingestion order, and publication identify different parts of that state. Together, they give backfill and feature evolution a stable read interface.

Future work extends aggregation over selected histories and overlaps historical prefetching with model computation. End-to-end model jobs can quantify these interactions within each training pipeline.

\section{Related Work}
\textbf{Recommendation and sequence infrastructure.}
Recommendation research spans collaborative filtering, feature engineering, and neural behavior modeling~\cite{resnick1997recommender,ricci2021recommender,bobadilla2013recommender,he2023survey,lai2025modeling}. Wide\&Deep, YouTube, and DIN established deep-model architectures~\cite{cheng2016wide,covington2016deep,zhou2018deep}; training systems co-design software and hardware for their access patterns~\cite{mudigere2022software}. Generative and sequence-oriented models demand longer histories and more computation~\cite{zhai2024actions,zhao2024recommender,chai2025longer,guan2025make,wang2025scaling,lin2026laser}, consistent with scaling laws~\cite{kaplan2020scaling}. Feature stores organize definitions, pipelines, and online/offline access~\cite{de2024hopsworks,liu2023optimizing,orr2021managing}. TokaDB unifies training-serving data management~\cite{tokadb}; Magnus integrates columnar ML data, versioning, and updates~\cite{song2025magnus}; Krypton combines real-time serving and analytical SQL~\cite{chen2023krypton}; ABase supports multi-tenant workloads~\cite{abase}. USDB couples incremental maintenance, shared histories, and inference-state reconstruction.

\textbf{Reconstruction and streaming state.}
Guo et al.~\cite{guo2026versioned} use versioned late materialization to separate reusable interaction history from training samples, with stream/batch visibility and projected prefetching. USDB's partial dumps retain a recent inference fragment and reconstruct the older prefix from version-compatible history. Fresh reads reuse the same maintenance and replicas. Lambda-style architectures and unified stream processors address complementary execution concerns~\cite{warren2015big,kreps2014questioning,zaharia2013discretized,carbone2015apache}; Flink 2.0 and Ursa extend state separation and lakehouse integration~\cite{mei2025flinkstate,merli2025ursa}. USDB coordinates user-history aggregation and inference visibility in shared state.

\textbf{LSM maintenance and access.}
LSM-trees and RocksDB supply the write-optimized foundation~\cite{o1996log,dong2021rocksdb}. Research explores compaction structures and workload-dependent tuning~\cite{dayan2017monkey,dayan2018dostoevsky,sarkar2022constructing,liu2024structural,huynh2022endure,yu2024camal,mo2023learning}; filters, learned indexes, and joins improve access within these structures~\cite{dayan2018optimal,luo2020rosetta,kraska2018case,dai2020wisckey,yu2024joins}. WiscKey and PebblesDB reorganize values or sorted runs to reduce storage work~\cite{lu2017wisckey,raju2017pebblesdb}, while AutoComp automates maintenance for lake tables~\cite{autocomp}. CascadeLSM places application-defined sequence aggregation at historical materialization boundaries. The resulting base is reused by serving and training.

\textbf{Distributed and disaggregated storage.}
Dynamo, Cassandra, Bigtable, and HBase establish distributed key-value and wide-table designs~\cite{decandia2007dynamo,lakshman2010cassandra,chang2008bigtable,vora2011hadoop}. Spanner, Aurora, and PolarDB explore distributed consistency and cloud resource separation~\cite{corbett2013spanner,verbitski2017amazon,cao2020polardb,polardb}. RocksDB disaggregation, CaaS-LSM, dLSM, LegoOS, and O3-LSM separate storage, compute, or maintenance duties~\cite{dong2023disaggregating,yu2024caas,wang2023dlsm,shan2018legoos,lin2026o3lsm}; GFS, HDFS, and Tectonic provide shared-file foundations~\cite{ghemawat2003google,borthakur2008hdfs,pan2021facebook}. USDB separates replica duties and reconciles shared history with fresh increments, complementing established consistency and availability contracts~\cite{brewer2000towards,gilbert2002brewer,abadi2012consistency,pritchett2008base}.

\textbf{Columnar layouts and encoding.}
PAX, C-Store, Vertica, and Parquet improve locality and selective access~\cite{ailamaki2001weaving,stonebraker2018c,lamb2012vertica,vohra2016apache}. BtrBlocks, Vortex, and Bullion develop adaptive encodings and formats for modern analytics or ML~\cite{kuschewski2023btrblocks,vortex2024,liao2025bullion}. F3 embeds portable decoders for extensible columnar representations~\cite{zeng2025f3}; OpenZL composes specialized codecs through a graph model and universal decoder~\cite{collet2025openzl}. USDB combines user indexing and columnar row groups for shared serving and projected training reads.

\section{Conclusion}
USDB shares materialized history across ingestion, inference, and training. Time-based compaction constructs reusable bases; differentiated replicas share their files; partial dumps reconstruct versioned inputs; and indexed columnar access accelerates selected histories. Production records show 85.8\% lower mean write latency and 72.6\% lower mean CPU with stable mean read latency; provisioning reduces normalized cost by 42.1\%. Across 374 controlled runs, repeated 10K/100K reads improve throughput 1.79$\times$/1.75$\times$, and longer 100K client-read jobs complete 1.62--2.52$\times$ faster. Sharing the sequence lifecycle supports both fresh serving and growing historical access.

\clearpage
\bibliographystyle{ACM-Reference-Format}
\bibliography{sample}

\clearpage
\end{document}